\newcommand{\PRformat}{} 
\ifdefined\PRformat
\documentclass[usenatbib, twocolumn, nofootinbib, superscriptaddress, longbibliography, reprint, prl]{revtex4-1}
\else
\documentclass[usenatbib, twocolumn, nofootinbib]{aastex631}
\fi

\usepackage{multirow}
\usepackage{bigints}
\usepackage{natbib}
\usepackage{color}
\usepackage{wrapfig}
\usepackage{amsmath,amsfonts,bm}
\usepackage{hyperref}
\usepackage[T1]{fontenc}
\usepackage{graphicx}   
\usepackage[mathlines]{lineno}
\usepackage{wrapfig}
\usepackage{fontawesome}
\usepackage{enumitem}
\usepackage{soul}

\newcommand{\prsectiontitleformat}[1]{\textbf{\emph{#1}}. ---}
\ifdefined\PRformat
\else
\newcolumntype{C}[1]{>{\centering\let\newline\\\arraybackslash\hspace{0pt}}m{#1}}
\fi

\ifdefined\PRformat
\definecolor{brown}{rgb}{0.8, 0.5, 0.2}
\fi

\newcommand{\sptpol}{SPTpol}
\newcommand{\sptthreeg}{SPT-3G}

\newcommand{\planck}{{\it Planck}}
\newcommand{\herschel}{{\it Herschel}}

\newcommand{\snr}{S/N}

\newcommand{\agora}{\textsc{Agora}}
\newcommand{\websky}{Websky}
\newcommand{\amber}{AMBER}

\newcommand{\taure}{\tau_{\rm re}}

\newcommand{\howmanysimulationsfornzero}{250}
\newcommand{\howmanysimulations}{100}

\newcommand{\degree}{\ensuremath{^\circ}}
\newcommand{\pixres}{0.^{\prime}5}
\newcommand{\ukam}{\ensuremath{\mu}{\rm K{\text -}arcmin}}
\newcommand{\sqdeg}{{\rm deg}^{2}}

\newcommand{\comment}[1]{}
\newcommand{\sz}{Sunyaev-Zel{'}dovich}

\newcommand{\zmid}{z_{\rm re}^{\rm mid}}
\newcommand{\zdur}{\Delta z_{\rm re, 50}}
\newcommand{\zduramber}{\Delta z_{\rm re, 90}}

\newcommand{\firstsptband}{95} 

\newcommand{\clcov}{{\bf C}_{\rm \ell}}
\newcommand{\clinv}{\clcov^{-1}}

\newcommand{\bigk}{\hat{K}}
\newcommand{\bigkmap}{\bigk({\bf \hat{n}})}
\newcommand{\clkkideal}{C_{L}^{KK}}
\newcommand{\clkk}{\hat{C}_{L}^{KK}}

\newcommand{\finalilckeynamelong}{minimum variance}
\newcommand{\finalilckeyname}{MV-LC}
\newcommand{\minLbigk}{50}
\newcommand{\maxLbigk}{300}
\newcommand{\nzerobias}{\hat{N}^{(0, KK)}_{L}}
\newcommand{\clkksys}{\hat{C}_{L,{\rm sys}}^{KK}}
\newcommand{\clphiphi}{C_{L, \kappa_{\rm CMB}}^{KK}}
\newcommand{\clphifg}{C_{L, \kappa_{\rm CMB}-FG}^{KK}}
\newcommand{\clfg}{C_{L, {\rm FG}}^{KK}}

\newcommand{\howmanygaussiansimulations}{250}

\newcommand{\fskyfinal}{f_{\rm sky, final}}
\newcommand{\fskylost}{f_{\rm sky, lost}}
\newcommand{\fieldsize}{\mbox{100 $\sqdeg$}}
\newcommand{\reionkszname}{reionization kSZ}

\newcommand{\elmin}{\ell_{\rm min}}
\newcommand{\elmax}{\ell_{\rm max}}
\newcommand{\elminvalue}{3300}
\newcommand{\elmaxvalue}{4300}
\newcommand{\amberboxsize}{1\ {\rm Gpc}/h}
\newcommand{\ambernparticles}{512}
\newcommand{\amberresolution}{2\ {\rm Mpc}}

\newcommand{\covsymbol}{\Sigma}
\newcommand{\covforlikelihoodfullinv}{\mathbf{\hat{\covsymbol}}_{L L^{\prime}}^{KK^{-1}}}
\newcommand{\covforlikelihoodfull}{\mathbf{\hat{\covsymbol}}_{L L^{\prime}}^{KK}}

\newcommand{\covforlikelihoodgau}{\mathbf{\hat{\covsymbol}}_{\rm Gau}^{KK}}

\newcommand{\ampcmbfg}{A_{\rm CMB-FG}}

\newcommand{\planckzmid}{7.69}
\newcommand{\planckzmiderror}{0.715}

\newcommand{\totalrawsnr}{10.3\sigma}

\newcommand{\zdurcombinedsimssnrcasea}{7.8\sigma}
\newcommand{\zdurcombinedsimssnrcaseb}{3.4\sigma}
\newcommand{\zdurcombinedsimsbestfitcasea}{3.9}
\newcommand{\zdurcombinedsimsbestfitcaseb}{4}
\newcommand{\zdurcombinedsimserrorcasea}{0.5}
\newcommand{\zdurcombinedsimserrorcaseb}{1.2}

\newcommand{\ampcmbfgbestfitvaluefromdatalowsys}{1.49 \pm 0.32}
\newcommand{\ampcmbfgbestfitvaluefromdata}{1.25 \pm 0.32}
\newcommand{\ampcmbfgbestfitvaluefromdatahighsys}{0.88 \pm 0.31}

\newcommand{\zdurtwosigmafromdataplanck}{13.0}
\newcommand{\zdurtwosigmafromdatakszplusGPzendatfive}{5}
\newcommand{\zdurtwosigmafromdatakszplusGPzendatsix}{4.5}
\newcommand{\zdurtwosigmafromdatakszplusGPplusplanckzendatsix}{2.6}

\newcommand{\zmidfromdatakszplusGPplusplanckzendatfive}{7.4 \pm 0.6}

\newcommand{\zdurcombinedsimscasebfortszoneptwosimcase}{4}
\newcommand{\zdurcombinedsimserrorcasebfortszoneptwpsimcase}{1.3}

\newcommand{\ptesimswithchisqworsethandata}{0.23} 
\newcommand{\ptefordatanull}{0.39} 

\newcommand{\zdurcombinedsimsellchoiceaunbiaseddeltaellfivehun}{3.8} 
\newcommand{\zdurcombinedsimsellchoiceaunbiaseddeltaellfivehunerr}{0.7}
\newcommand{\zdurcombinedsimsellchoiceaunbiaseddeltaellsevenhun}{3.8} 
\newcommand{\zdurcombinedsimsellchoiceaunbiaseddeltaellsevenhunerr}{0.6}
\newcommand{\zdurcombinedsimsellchoiceaunbiased}{3.9}
\newcommand{\zdurcombinedsimsellchoiceaunbiasederr}{0.5}

\newcommand{\abstracttext}
{
We report results from an analysis aimed at detecting the trispectrum of the kinematic \sz{} (kSZ) effect by combining data from the South Pole Telescope (SPT) and \herschel-SPIRE experiments over a \fieldsize{} field. 
The SPT observations combine data from the previous and current surveys, namely \sptpol{} and \sptthreeg, to achieve depths of 4.5, 3, and 16 $\ukam$ in bands centered at \firstsptband, 150, and 220 GHz. 
For SPIRE, we include data from the 600 and 857 GHz bands. 
We reconstruct 
the velocity-induced large-scale correlation of the small-scale kSZ signal with a quadratic estimator that uses two cosmic microwave background (CMB) temperature maps, constructed by optimally combining data from all the frequency bands.
We reject the null hypothesis of a zero trispectrum at $\totalrawsnr$ level. 
However, the measured trispectrum contains contributions from both the kSZ and other undesired components, such as CMB lensing and astrophysical foregrounds, with kSZ being sub-dominant.
We use the \agora{} simulations to estimate the expected signal from CMB lensing and astrophysical foregrounds.
After accounting for the contributions from CMB lensing and foreground signals, we do not detect an excess kSZ-only trispectrum and use this non-detection to set constraints on reionization.
By applying a prior based on observations of the Gunn-Peterson trough, we obtain an upper limit on the duration of reionization of $\zdur < \zdurtwosigmafromdatakszplusGPzendatsix$ (95\% C.L).
We find these constraints are fairly robust to foregrounds assumptions. 
This trispectrum measurement is independent of, but consistent with, \planck's optical depth measurement.
This result is the first constraint on the epoch of reionization using the non-Gaussian nature of the kSZ signal.
}

\newcommand{\tittext}{First Constraints on the Epoch of Reionization Using the non-Gaussianity of the Kinematic \sz{} Effect from the South Pole Telescope and \bf{\emph{Herschel}}-SPIRE Observations}

\ifdefined\PRformat
\newcommand{\mnras}{MNRAS}
\newcommand{\aap}{AAP}
\newcommand{\procspie}{SPIE}
\newcommand{\apjs}{ApJS}
\newcommand{\apjl}{ApJL}
\newcommand{\jcap}{JCAP}
\newcommand{\pasp}{PASP}
\newcommand{\araa}{ARA\&A}
\fi

\defcitealias{smith17}{SF17}
\defcitealias{ferraro18}{FS18}
\defcitealias{raghunathan23}{RO23}

\begin{document}

\title{\tittext}
\newcommand{\shortauthourlist}{S. Raghunathan}

\ifdefined\PRformat
\else
\shorttitle{Probing Reionization Using {\rm kSZ} Trispectrum}
\shortauthors{\shortauthourlist}
\fi

\author{S.~Raghunathan}
\affiliation{Center for AstroPhysical Surveys, National Center for Supercomputing Applications, Urbana, IL, 61801, USA}
\author{P.~A.~R.~Ade} \affiliation{School of Physics and Astronomy, Cardiff University, Cardiff CF24 3YB, United Kingdom}
\author{A.~J.~Anderson} \affiliation{Fermi National Accelerator Laboratory, MS209, P.O. Box 500, Batavia, IL, 60510, USA} \affiliation{Kavli Institute for Cosmological Physics, University of Chicago, 5640 South Ellis Avenue, Chicago, IL, 60637, USA} \affiliation{Department of Astronomy and Astrophysics, University of Chicago, 5640 South Ellis Avenue, Chicago, IL, 60637, USA}
\author{B.~Ansarinejad} \affiliation{School of Physics, University of Melbourne, Parkville, VIC 3010, Australia}
\author{M.~Archipley} \affiliation{Department of Astronomy, University of Illinois Urbana-Champaign, 1002 West Green Street, Urbana, IL, 61801, USA} \affiliation{Center for AstroPhysical Surveys, National Center for Supercomputing Applications, Urbana, IL, 61801, USA}
\author{J.~E.~Austermann} \affiliation{NIST Quantum Devices Group, 325 Broadway Mailcode 817.03, Boulder, CO, 80305, USA} \affiliation{Department of Physics, University of Colorado, Boulder, CO, 80309, USA}
\author{L.~Balkenhol} \affiliation{Institut d'Astrophysique de Paris, UMR 7095, CNRS \& Sorbonne Universit\'{e}, 98 bis boulevard Arago, 75014 Paris, France}
\author{J.~A.~Beall} \affiliation{NIST Quantum Devices Group, 325 Broadway Mailcode 817.03, Boulder, CO, 80305, USA}
\author{K.~Benabed} \affiliation{Institut d'Astrophysique de Paris, UMR 7095, CNRS \& Sorbonne Universit\'{e}, 98 bis boulevard Arago, 75014 Paris, France}
\author{A.~N.~Bender} \affiliation{High-Energy Physics Division, Argonne National Laboratory, 9700 South Cass Avenue, Lemont, IL, 60439, USA} \affiliation{Kavli Institute for Cosmological Physics, University of Chicago, 5640 South Ellis Avenue, Chicago, IL, 60637, USA} \affiliation{Department of Astronomy and Astrophysics, University of Chicago, 5640 South Ellis Avenue, Chicago, IL, 60637, USA}
% ...
 \collaboration{SPT-3G and SPTpol Collaboration} \noaffiliation

\begin{abstract}
\abstracttext{}
\end{abstract}

\ifdefined\PRformat
\maketitle
\fi



\ifdefined\PRformat
\prsectiontitleformat{Introduction}
\else
\section{Introduction}
\label{sec_introduction}
\fi
The kinematic \sz{} (kSZ) effect originates when electrons with bulk motion Compton-scatter cosmic microwave background (CMB) photons \citep{sunyaev80b}. 
Detecting the kSZ signal can provide crucial insights on both structure formation \citep{mueller15a, bianchini16} and the physics of reionization \citep{knox98, mcquinn05, zahn12, reichardt15, choudhury22, jain23}. 
This is because the source of the kSZ signal can be decomposed into two main categories: a low-redshift ($z \lesssim 3$) component referred to as homogeneous- or post-reionization kSZ and a high-redshift ($z \gtrsim 6$) component referred to as inhomogeneous or reionization-kSZ. 
The post-reionization kSZ signal is due to the bulk flow of halos with free electrons in the local Universe. 
The \reionkszname, on the other hand, is due to motion of the ionized bubbles containing free electrons during the epoch of reionization (EoR). 
Several observations \citep[e.g,][]{becker01, fan06, robertson10} suggest that the energetic ultraviolet light from the first stars and galaxies at $z \gtrsim 6$ was responsible for ionizing the neutral hydrogen in the early Universe---although more data is required to precisely understand the process, timing, and duration of the EoR \citep{robertson22}.
From the CMB data, besides kSZ, the EoR can also be probed using the large-scale bump in the CMB $EE/TE$ power spectra \citep{zaldarriaga97a}. 
Measurements of the large-scale ($\ell \lesssim 10$) modes from ground-based experiments, because of the limited sky coverage, will be dominated by the sample variance which reduces the sensitivity to the low-$\ell$ EE/TE reionization bump. In addition to this, the low-$\ell$ bump is not sensitive to the details of the process of reionization and cannot distinguish between different reionization histories \citep[see Fig. 4 of][]{planck16-48}. Hence, it is important to explore other probes of EoR, like the kSZ.

The kSZ signals are sub-dominant compared to other signals in maps of total intensity 
and detecting them has proved to be challenging. 
In the past, the kSZ signals have been detected through cross-correlation of CMB maps with galaxy surveys \ifdefined\PRformat
\citep{hand12, hill16, calafut21, schaan21, schiappucci23, mallaby-kay23}
\else
\citep[for example:][]{hand12, hill16, calafut21, schaan21, schiappucci23, mallaby-kay23}
\fi
but these measurements only probe the post-reionization kSZ signal. 
Although cross-correlations with high redshift galaxy catalogs are in principle possible for the \reionkszname, the expected signal-to-noise ($\snr$) is small ($2-3\sigma$) even for future Stage-4 experiments because of the difficulty in obtaining galaxy catalogues at $z \gtrsim 6$ \citep{laplante22}. 
Hence, forecasts for kSZ constraints on the EoR have typically relied on kSZ power spectrum (2-pt function) measurements or reconstructions of the optical depth \citep{dvorkin09}.
The kSZ 2-pt function, however, receives  contributions from both of the kSZ components and disentangling the two is difficult due to their similar shapes and amplitudes \citep{shaw12, battaglia13}. 
The presence of astrophysical foregrounds in the CMB maps, especially thermal SZ (tSZ) and cosmic infrared background (CIB) signals, complicates the interpretation of the kSZ power spectrum further as these foreground signals are much brighter than kSZ \citep[][hereafter \citetalias{raghunathan23}]{raghunathan23}. 
As a result, EoR constraints from recent measurements of the total kSZ power spectrum \citep{reichardt21, gorce22} have been limited by knowledge of the foreground and post-reionization kSZ signal.

\citet[hereafter \citetalias{smith17}]{smith17} proposed a novel method of using the kSZ trispectrum (4-pt function) to probe the physics of reionization \citep[also see][]{ferraro18, alvarez20}. 
Since the \reionkszname{} signal depends on both the free electron density and the velocity, the small-scale fluctuations in electron density due to inhomogeneous reionization will get modulated on larger scales by the velocity field, leading to position-dependent non-Gaussianities in the CMB maps. 
Thus, the kSZ non-Gaussianity arises due to the large scale correlations of small scale clustering of halos, similar to CMB lensing \citepalias{smith17}. 
However, in our case the large scale correlations are due to the bulk velocity flow in the Universe rather than due to gravitational lensing \citep{okamoto03}. 
Although the post-reionization kSZ also starts out as a non-Gaussian signal at different epochs, 
the comoving line-of-sight distance over which it gets integrated in the local Universe is much larger and the signal ends up being Gaussian based on the central limit theorem. 
This characteristic allows the two kSZ components to be easily distinguished using the trispectrum \citepalias[see Fig.~2 of][]{smith17}. 
Despite this advantage, the constraints on reionization from kSZ 4-pt alone are not expected to be as competitive as the optical depth measurements from \planck{} even for future CMB surveys, due to the degeneracy between the parameters that govern reionization. However, as demonstrated by \citet{alvarez20}, the joint constraints from kSZ 4-pt, \planck{} primary CMB, and kSZ 2-pt can effectively break that degeneracy, resulting in a significant improvement on reionization compared to what can be achieved individually by any of these probes.

In this work, we present results from an analysis aimed at detecting the kSZ trispectrum using CMB temperature maps obtained by combining South Pole Telescope (SPT) and \herschel-SPIRE datasets. 
The observed trispectrum receives a contribution from \reionkszname{} but is dominated by CMB lensing and astrophysical foreground signals. 
We build a template for the latter using the \agora{} simulations \citep{omori22}. 
Given the difficulties in correctly modeling the foreground signals and the lower amplitude of the kSZ signals compared to the other undesired signals, we adopt different strategies to handle the foregrounds. 
In the baseline case, we marginalize over the amplitude of the CMB lensing and foreground signals. 
We also take the approach of fixing the CMB lensing and foreground signals. 
In neither case do we observe an excess kSZ trispectrum. 
We use this non-detection along with a prior based on the measurements of the Gunn-Peterson (GP) trough to set upper limits (95\% C.L.) on the duration of reionization $\zdur$ corresponding to the difference in redshifts at which the Universe has been 25\% and 75\% reionized. We show that our results are consistent with \planck's optical depth measurement. 
We do not combine our results with the kSZ 2-pt measurements from the literature \citep{reichardt21, gorce22} owing to assumptions made about foregrounds and the post-reionization kSZ signals made in those works.\ifdefined\PRformat\\\fi


\ifdefined\PRformat
\prsectiontitleformat{CMB maps}
\else
\section{Datasets}
\label{sec_datasets}
\fi
This work uses data from two different experiments: SPT \citep{padin08, carlstrom11} and \herschel-SPIRE \citep{pilbratt10, griffin10}. 
For SPT, we use data from two surveys: \sptpol{} \citep{austermann12} and \sptthreeg{} \citep{benson14, bender18, sobrin22}. 
The \sptpol{} observations were carried out between 2012 and 2016, and in this work we only use the 150 GHz observations \citep{henning18} since the noise level of \firstsptband{} GHz \sptpol{} is roughly $\times3$ higher than the equivalent \sptthreeg{} data. 
The \sptthreeg{} observations used in this work were carried out between 2019 and 2020, and we include data from all the three bands: \firstsptband, 150, and 220 GHz.  
After combining \sptpol{} and \sptthreeg, the map depths for the three bands are: 4.5, 3, and 16 \ukam{} respectively. 
For \herschel-SPIRE, we use the data from 600 GHz (500 $\mu$m) and 857 GHz (350 $\mu$m) bands. 
Since \herschel-SPIRE is primarily used for CIB mitigation and since the CIB-decorrelation between SPT bands and \herschel-SPIRE's 1200 GHz (250 $\mu m$) band is high \citep{viero19}, we do not use the 1200 GHz band in this work.
We limit the SPT footprint in this work to the region that has overlap with \herschel-SPIRE, which is a roughly \fieldsize{} region centered at \mbox{(RA, Decl.) = (23h30m, -55$\degree$)}. We provide details about the data processing in \ifdefined\PRformat the Appendix. \else Appendix~\ref{sec_data_processing}.\fi In short, the raw SPT data is filtered and binned into maps with a pixel resolution of $\pixres$. The effect of filtering is accounted for by using the transfer function (TF) calculated from simulations. The individual frequency maps are then calibrated by cross-correlating with \planck. 
The SPIRE and calibrated SPT maps are combined to produce a minimum-variance (MV) map which has an unbiased response to CMB temperature. 
This is done using a scale-dependent linear combination technique \citep{cardoso08}, and we refer to the product as the \finalilckeyname.
\ifdefined\PRformat
\\
\fi


\ifdefined\PRformat
\prsectiontitleformat{\bm{$\bigkmap$} and \bm{$\clkk$} measurements}
\else
\section{Results and Discussion}
\label{sec_results_and_discussion}
\subsection{$\bigkmap$ and $\clkk$ measurements}
\label{sec_clkk_measurement}
\fi
We briefly describe the methods to extract \reionkszname{} information from the CMB maps and refer the reader to Appendix B for more details. Following the work of \citetalias{smith17}, we develop a quadratic estimator (QE) to reconstruct $\bigkmap$ which captures the degree-scale correlations of small-scale clustering of the kSZ signal. 
The desired kSZ trispectrum \mbox{$\clkk = \delta(L-L^{\prime}) \bigk_{L} \bigk_{L^{\prime}}^{\ast}$} is the power spectrum of the reconstructed $\bigkmap$ map \citepalias{smith17} after removing an estimate of the mean-field which arises due to \ifdefined\PRformat masking\else masking (see \S{\ref{sec_N0_meanfield_covariance_estimation}}\fi. 
For $\clkk$ measurement, we set a bin width of $\Delta_{L} = 50$ in the range $L \in [50, 300]$. 
Besides the desired kSZ signal and the mean-field, the reconstructed map receives contributions from the following: $\nzerobias$ which is the Gaussian disconnected piece arising due to chance correlations of the two CMB maps used in the QE; and the systematics from CMB lensing and foregrounds. We use \amber{} (\agora) simulations to model the \reionkszname{} (CMB lensing and foregrounds) and assume a Gaussian likelihood to derive constraints on reionization.

\begin{figure}
\centering
\includegraphics[width=0.5\textwidth, keepaspectratio]{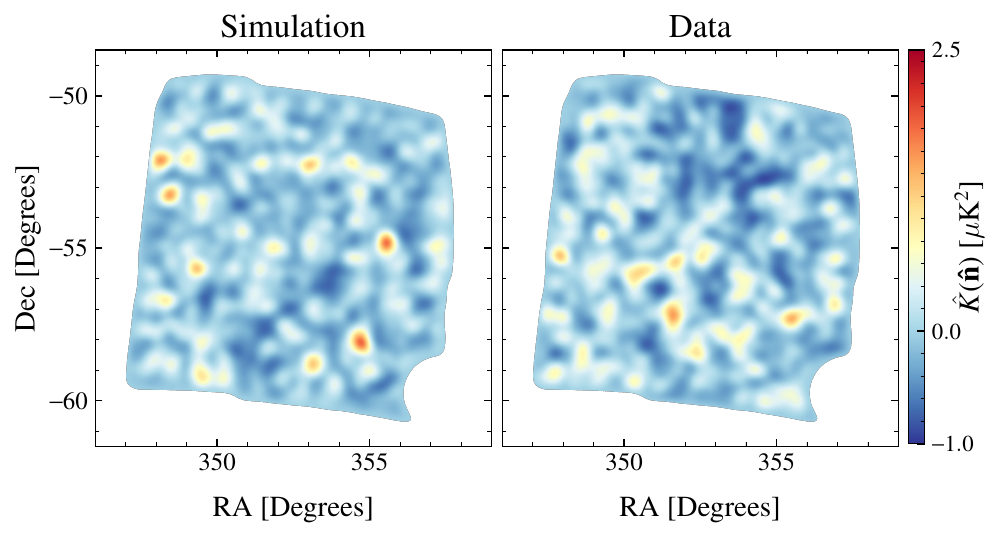}
\caption{Mean-field subtracted $\bigkmap$ maps from a single simulation run in the left panel and data in the right panel. 
Both the panels have been smoothed using a Gaussian beam with $\theta_{\rm FWHM} = 30^{\prime}$.
The figure illustrates that our data and simulations have qualitatively similar-looking features. 
The simulation only contains CMB, foregrounds, and noise. It does not include the \reionkszname{} signal. The similarity of the simulation with the data map suggests that the reconstructed $\bigkmap$ is dominated by $\nzerobias$ and foregrounds. 
We also provide quantitative comparison between the data and the simulations in the text.
}
\label{fig_bigk_maps_data_sims}
\end{figure}

The left panel of Fig.~\ref{fig_bigk_maps_data_sims} shows the $\bigkmap$ reconstructed from a single non-Gaussian simulation run while the right panel is for data. 
We note that the statistical properties look qualitatively similar between the panels.
The simulation includes CMB, foregrounds, and noise but does not contain the \reionkszname{} signal. 
The similarity between the simulation and data suggests that our reconstructed $\bigkmap$ should be dominated by $\nzerobias$ and foregrounds, and the kSZ signal must be sub-dominant.

\begin{figure}
\centering
\includegraphics[width=0.48\textwidth, keepaspectratio]{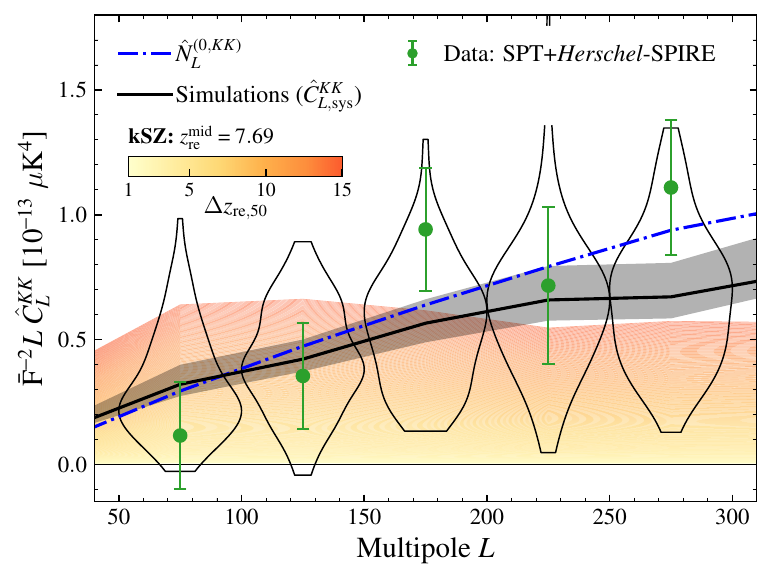}
\caption{Reconstructed $\clkk$ signals: 
The violins represent the scatter in the measurements from the \howmanysimulations{} non-Gaussian (\reionkszname-free) simulations in each $L$ bin and green data points represent data.
The mean of all the simulations ($\clkksys$) is shown in black. 
The band around the black curve indicates the systematic uncertainty in the template obtained by scaling the tSZ signal in the input simulations by $\pm 20 \%$.
For each curve, the estimate of $\nzerobias$ (blue dash-dotted curve) has been removed. 
For reference, we also show the expected kSZ signal from \amber{} simulations for $\zmid = \planckzmid$ and different values of $\zdur$ in shades of orange. 
From the figure, it is evident that the data is consistent with the non-Gaussian simulations, which do not contain any \reionkszname{} signal, in agreement with the $\bigkmap$ maps in Fig.~\ref{fig_bigk_maps_data_sims}. 
}
\label{fig_clkk_data_sims}
\end{figure}

\begin{figure}
\centering
\includegraphics[width=0.48\textwidth, keepaspectratio]{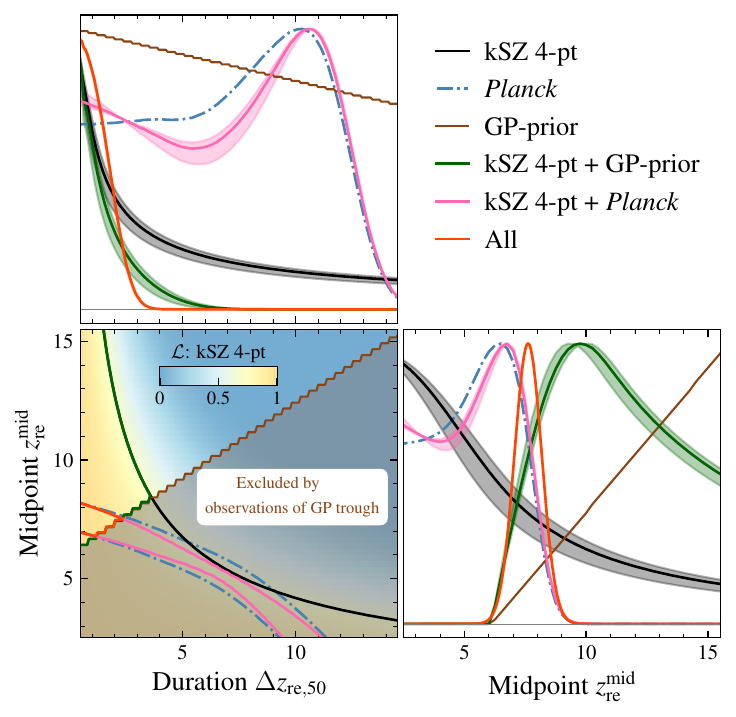}
\caption{
Constraints on EoR parameters ($\zdur$ and $\zmid$) for different dataset combinations after marginalising over $\ampcmbfg$: We present both 2D and 1D posteriors. The two parameters are highly degenerate for kSZ 4-pt alone, as shown in the 2D posterior plot. The black line represents the 68\% C.L. joint upper limit to the two parameters in the 2D plot 
and the marginalized 1-parameter posteriors in the 1D plots. The green lines show the analogous curves after applying a prior based on GP trough measurements which removes regions where reionization ends later than $z = 6$. 
For reference, the GP-prior is shown in brown.
We adopt these as our baseline constraints, and for this case we set a 95\% C.L. upper limit of \mbox{$\zdur < \zdurtwosigmafromdatakszplusGPzendatsix$}. 
We also show 68\% C.L. allowed 2D regions and 1D posterior curves derived from \planck's low-$\ell$ optical depth measurement (blue dash-dotted curve) and from combining \planck{} and kSZ 4-pt with (without) the GP-based prior in red (pink). 
The band around the combinations that include kSZ 4-pt data represents the uncertainties in $\clkksys$ template. To avoid cluttering, we show the band only in the 1d posteriors. 
}
\label{fig_posterior_data}
\end{figure}

In Fig.~\ref{fig_clkk_data_sims} we present the $\clkk$ measurements from simulations and data. 
These correspond to the results with our fiducial filter choices \mbox{$(\elmin, \elmax) = (\elminvalue, \elmaxvalue)$} in Eq.(\ref{eq_final_filter}). 
The result from data (green circles) is consistent with the distribution of the simulations represented by the violins. 
With our data, we reject the null hypothesis of a zero trispectrum at $\totalrawsnr$. 
This raw $\snr$ has been calculated just with the Gaussian covariance $\covforlikelihoodgau$ in Eq.(\ref{eq_likelihood}).
The mean of all the simulations, which we use as the estimate of $\clkksys$ is the black solid curve. 
The semi-transparent band around the mean is the systematic uncertainty in $\clkksys$ obtained by scaling the tSZ signal in the input simulations by $\pm 20 \%$ roughly consistent with the uncertainty in the  hydrostatic mass bias parameter \citep{planck15-24}. The impact of this systematic on our constraints is presented in Fig.~\ref{fig_posterior_data} and Fig.~\ref{fig_posterior_data_ksz_4pt_only}.
The change in the results is only marginal if we scaled the CIB instead of the tSZ.
The blue dash-dotted curve is $\nzerobias$ calculated using \howmanygaussiansimulations{} simulations, and it has been removed from all the other curves in the figure. 
The error bars include contribution both from the scatter in the Gaussian $\nzerobias$ and the non-Gaussian signals. 
For reference, we also show the expected kSZ 4-pt function signal in shades of orange. 
These assume a fixed midpoint $\zmid = \planckzmid$ and different values of duration $\zdur \in [1, 15]$. 

The probability to exceed (PTE or $p$-value), obtained by comparing the individual simulations (distributions shown using the violins) and data (green) with the $\clkksys$ template (black) and computing \mbox{$p = \chi^{2}_{\rm sims} \ge \chi^2_{\rm data}$} is \mbox{$p = \ptesimswithchisqworsethandata$} 
indicating the consistency between the data and the simulations. 
Note that, similar to Fig.~\ref{fig_bigk_maps_data_sims}, the simulations do not include the \reionkszname{} signal. 
This suggests that the \reionkszname{} signal must be sub-dominant compared to $\nzerobias$ and 4-pt function contributions from CMB lensing and foregrounds $\clkksys$. 
Indeed, when we remove the $\clkksys$ estimate from data, the residual $\clkk$ measurement is consistent with a null signal with \mbox{$p = \ptefordatanull$}.
\ifdefined\PRformat
\\
\fi

\ifdefined\PRformat
\prsectiontitleformat{Reionization constraints}
\else
\subsection{Reionization constraints}
\label{sec_delta_z_constraints}
\fi
We compare the $\clkk$ measurements obtained above to the expected kSZ signal from \amber{} to place constraints on the EoR parameters. 
We fit for three parameters $[\zmid$, $\zdur$, and $\ampcmbfg]$ where $\ampcmbfg$ is the amplitude term for the \agora{} $\clkksys$ template. \ifdefined\PRformat In the Appendix, \else Appendix~\ref{sec_pipeline_validation},\fi we use simulations to show that this approach is robust to the assumptions about the $\clkksys$ template and returns unbiased results on simulated data. 

In Fig.~\ref{fig_posterior_data}, we present the constraints on $\zdur$ and $\zmid$ for different dataset combinations. 
As evident from the figure, the two parameters are degenerate for the kSZ 4-pt-only case shown in the 2D posterior. 
On the other hand, evidence based on the measurements of GP trough \citep{becker01, fan06} in the spectra of high redshift quasars suggest that hydrogen in the Universe must be fully ionized by $z \sim 5-6$. 
We use this information to set a binary GP-based prior
(brown)
on the reionization histories $X_{\rm e}(z)$ from \amber. 
To this end, we remove regions in the $(\zmid-\zdur)$ plane where reionization ends later than $z < 6$ by setting their prior to zero. 
We define the end of reionization as the redshift at which $X_{\rm e}(z) > 0.95$. 
The kSZ 4-pt result combined with the GP-based prior is shown in green, and we adopt this as our baseline result.  
For this case, we are able to set an upper limit on \mbox{$\zdur$ of $< \zdurtwosigmafromdatakszplusGPzendatsix$ (95\% C.L.)}.
Modifying the GP-based prior to have the reionization end at $z < 5$ rather than $z < 6$, slightly increases the upper limit to \mbox{$\zdur < \zdurtwosigmafromdatakszplusGPzendatfive$ (95\% C.L.)}.

We also show constraints based on the \planck{} optical depth measurement (blue dash-dotted curve) as well as the kSZ 4-pt measurement + \planck{} with and without the GP-based prior (pink and red curves respectively) in Fig.~\ref{fig_posterior_data}.
As expected, \planck{} (blue) tightly constrains the upper tail of the midpoint of reionization but is not sensitive to the duration with an upper limit of \mbox{$\zdur$ of $< \zdurtwosigmafromdataplanck$ (95\% C.L.)}.
Adding kSZ 4-pt to \planck{} without the GP-based prior (pink) leads to marginal ($<10\%$) changes in the results. 
On the other hand, by including the GP-based prior to kSZ 4-pt + \planck{} (red), we set an upper limit on \mbox{$\zdur$ of $< \zdurtwosigmafromdatakszplusGPplusplanckzendatsix$ (95\% C.L.)} and obtain \mbox{$\zmid = \zmidfromdatakszplusGPplusplanckzendatfive$.} 
This combination is dominated by \planck{} and the GP-based prior, and kSZ 4-pt only adds marginal improvement, however, and we do not quote it as our main result.

For the amplitude of the CMB lensing and foreground template, we obtain a best-fit amplitude of \mbox{$\ampcmbfg = \ampcmbfgbestfitvaluefromdata$}.
The detection significance for $\ampcmbfg$ is consistent with what we expect from simulations shown in the inset plots of Fig.~\ref{fig_delta_z_posterior_sim_tests}.
As shown in Fig.~\ref{fig_posterior_data_ksz_4pt_only}, we do not observe a strong degeneracy between $\ampcmbfg$ and the EoR parameters.

To summarise, we quote the constraint from kSZ 4-pt with the GP-based prior (green curves in Fig.~\ref{fig_posterior_data}) as our baseline result and we set an upper limit on the duration of reionization of \mbox{$\zdur < \zdurtwosigmafromdatakszplusGPzendatsix$ (95\% C.L.)}.
These are the first constraints on EoR parameters using the non-Gaussianity of the kSZ signal.
\ifdefined\PRformat
\\
\fi

\ifdefined\PRformat
\prsectiontitleformat{Foreground uncertainties}
\else
\subsubsection{Foreground uncertainties}
\label{sec_delta_z_constraints}
\fi
In Fig.~\ref{fig_posterior_data}, the band around the dataset combinations that include kSZ 4-pt data represents the uncertainties in $\clkksys$ propagated to uncertainties in the parameter constraints. 
These are obtained by scaling the tSZ signal in the simulations by 20\% (semi-transparent black band around $\clkksys$ in Fig.~\ref{fig_clkk_data_sims}). 
As evident from the figure, the uncertainties in $\clkksys$ have negligible effect indicating constraints are robust to assumptions about the CMB+FG template. For example, the 95\% C.L. of $\zdur$ changes by $\sim7\%$ when the $\clkksys$ is modified to take the uncertainties into account.
The above results are for our fiducial values for \mbox{$(\elmin, \elmax) = (\elminvalue, \elmaxvalue)$} with $\Delta_{\ell} = 1000$. 
We find these values to be optimal both in terms of $\snr$ and the foreground biases.
\ifdefined\PRformat In the Appendix, \else In Appendix~\ref{sec_bias_cmb_fg},\fi we justify this by discussing the changes to $\snr$ and the impact of foregrounds when the values of $\elmin, \elmax, \Delta_{\ell}$ are modified. 
We also show that the systematics due to uncertainties in beam and TF to be negligible. 
\ifdefined\PRformat
\\
\fi

\ifdefined\PRformat
\prsectiontitleformat{Conclusion}
\else
\section{Conclusion}
\label{sec_conclusion}
\fi
In this work, we reported results from an analysis aimed at detecting the non-Gaussian nature of the kSZ signal from reionization. 
We combined data from \sptpol, \sptthreeg, and {\it Herschel}-SPIRE surveys in a \fieldsize{} field. 
Using \finalilckeyname{} map and a QE, we detected the total trispectrum $\clkk$ at $\totalrawsnr$. 
The measurement is dominated by $\nzerobias$ bias from the disconnected term, and the contributions from CMB lensing and foreground signals. 
After accounting for these undesired signals, we do not measure an excess kSZ trispectrum and our results are consistent with a null signal ($p = \ptefordatanull$). 
The results from data are consistent with the expectations from simulations ($p=\ptesimswithchisqworsethandata$). 
We quantified the biases due to uncertainties in instrumental beam and TF, and found them to be negligible. 
We also thoroughly checked 
the biases due to mismatch between the \agora{} foreground template and data, and found that our results are robust to uncertainties in both amplitude and shape of the template. 


We found that the constraints on $\zmid$ and $\zdur$ are highly degenerate from kSZ 4-pt alone. 
Hence, we applied a loose prior based on GP trough measurements on the reionization histories from \amber{} to remove $\zmid$ and $\zdur$ from the parameter space where the reionization ends later than $z_{\rm end} = 6$. 
With this prior, we set a upper limit (95\% C.L.) of $\zdur < \zdurtwosigmafromdatakszplusGPzendatsix$. 
This result is independent of, but consistent with, the optical depth measurement from \planck. 
This work represents the first constraints on EoR parameters using the trispectrum of the kSZ signal.
In the current work, we have used simulations to model the contributions from lensing and foregrounds. There are other potential strategies to mitigate them at the expense of a penalty in $\snr$. These include bias-hardening \citep{namikawa13, osborne14, sailer21} and delensing \citep{manzotti17, ade21_bicepdelensing}, which might be particularly relevant for the future surveys and should be explored further. Furthermore, we do not include information from kSZ power spectrum here, owing to the uncertainties in the contribution from the post-reionization kSZ signal to the total kSZ power spectrum. In the future, however, our understanding about the post-reionization kSZ signal is expected to improve, thanks to the synergies between CMB and galaxy surveys. Another potential approach is to use the cross-correlation between CMB and galaxy surveys for ``de-kSZing'' the post-reionization kSZ signal \citep{foreman23}.

Besides the kSZ measurements, the high redshift measurements of quasars \citep[recently,][]{dayal24, munoz24} from JWST \citep[for a review, see][]{robertson22}, the upcoming 21cm measurements from experiments like HERA and SKA \citep{liu20, HERA22}, and the cross-correlation between multiple probes \citep[for example,][]{namikawa21, laplante23, georgiev23} are all expected to significantly enhance our understanding of the physics of EoR in the next decade. 
The kSZ results are also expected to improve with the upcoming low-noise multi-frequency CMB datasets; namely from the full-depth \sptthreeg{} and \sptthreeg+ \citep{benson14, bender18, anderson22}, Simons Observatory \citep{simonsobservatorycollab19}, 
CCAT \citep{cothard20}
 and CMB-S4 \citep{abazajian19} surveys, which all cover much wider sky area compared to this work. 
Interpretation of results from the future measurements will necessitate the development of more kSZ simulations like 21cmFAST \citep{mesinger10} and \amber{} \citep{trac22}; and also new and multiple realizations of the correlated extragalactic skies like \agora{} \citep{omori22}, Sehgal \citep{sehgal10}, and \websky{} \cite{stein20}.
The combination of kSZ 4-pt function measurements from the upcoming surveys along with high-significance measurements of the kSZ power spectrum \citepalias{raghunathan23} and optical depth measurements from LiteBIRD \citep{litebird23} can help to break degeneracies between $\zmid$ and $\zdur$ \citep{alvarez20} forming a powerful probe of the epoch of reionization. 
This work forms a key first step towards such future high precision measurements. 

The data files and plotting scripts used in this work are available from this \href{https://github.com/sriniraghunathan/kSZ_4pt_SPT_SPIRE}{link$^{\text{\faGithub}}$}.

\section*{Acknowledgments}
We thank all the three anonymous referees for their detailed and valuable feedback which has helped in shaping the manuscript better.
SR acknowledges support by the Illinois Survey Science Fellowship from the Center for AstroPhysical Surveys at the National Center for Supercomputing Applications.

This work made use of the following computing resources: Illinois Campus Cluster, a computing resource that is operated by the Illinois Campus Cluster Program (ICCP) in conjunction with the National Center for Supercomputing Applications (NCSA) and which is supported by funds from the University of Illinois at Urbana-Champaign; the computational and storage services associated with the Hoffman2 Shared Cluster provided by UCLA Institute for Digital Research and Education's Research Technology Group; and the computing resources provided on Crossover, a high-performance computing cluster operated by the Laboratory Computing Resource Center at Argonne National Laboratory.

The South Pole Telescope program is supported by the National Science Foundation (NSF) through award OPP-1852617. Partial support is also provided by the Kavli Institute of Cosmological Physics at the University of Chicago.

Work at Argonne National Lab is supported by UChicago Argonne LLC, Operator of Argonne National Laboratory (Argonne). Argonne, a U.S. Department of Energy Office of Science Laboratory, is operated under contract no. DE-AC02-06CH11357.
\ifdefined\PRformat
\appendix
\section{APPENDIX}
\else
\appendix
\fi

\ifdefined\PRformat
\section{A. Data processing}
\label{sec_data_processing}
\subsection{A.1. Map making}
\label{subsubsec_mapmaking}
\else
\section{Data processing}
\label{sec_data_processing}
\subsection{Map making}
\label{subsubsec_mapmaking}
\fi
The mapmaking procedure for both the \sptpol{} and \sptthreeg{} surveys is similar to previous SPT works \citep{schaffer11, henning18, dutcher21} and we direct the readers to those works for more details. 
Briefly, the time-ordered data (TOD) from each detector are binned using a flat-sky approximation in the Sanson-Flamsteed projection \citep{calabretta02, schaffer11} with a pixel resolution of $\pixres$. 
We employ the following filtering schemes to the TOD before binning them into maps.
First is the removal of the common-mode, which corresponds to the mean signal from detectors in a given band. 
Next, to remove excess noise in the scan direction $\ell_{x}$, we fit for and remove a combination of Legendre polynomials (up to $7^{th}$ order) and sines and cosines (up to an effective high-pass cutoff of $\ell_x = 300$.  
Finally, to prevent the aliasing in the step of binning into map pixels, we low-pass filter the TOD at a frequency equivalent to $\ell_x = 13000$.
The \herschel-SPIRE maps used in this work are the same as those used in previous SPT works \citep{holder13, viero19} and we refer the reader to those works for more details. 
\newcommand{\tftwod}{{\bf F_{\ell}}}
We calculate the effect of the above TOD filtering using end-to-end simulations. 
We make Gaussian realizations of an underlying power spectrum and mock-observe them with our map-making pipeline. 
The ratio of the power spectrum of the output maps over the input maps -- filter transfer function (TF) -- captures the effect of filtering. 
To account for anisotropic filtering in SPT data, 
we compute the TF in the 2D as a function of $\ell_{x}$ and $\ell_{y}$. 
We denote the azimuthal average of the 2D TF as $F_{\ell}$.
We use 250 mock observations of a white noise power spectrum with $\Delta_{T} = 10 \ukam$ to estimate the TF. 
We do not note significant differences in TF when we replace the white noise mocks with CMB and foregrounds realizations. 
The TF is slightly different at low $\ell$ for \sptpol{} and \sptthreeg{} because of the size of the focal-plane,
but the $F_{\ell}$ is close to 95\% and roughly flat for all the SPT bands in the range $\ell \in [3000, 5000]$, which is the $\ell$ range most relevant to this analysis. 
For \herschel-SPIRE data, a high-pass filter along both $|\ell_{x}| \lesssim 200$ and $|\ell_{y}| \lesssim 200$ has been applied to remove excess large-scale noise. 
This, however, has negligible impact on $F_{\ell}$ which is unity in our desired $\ell$ range \citep[See Fig. 2 of][]{viero13a}. 
The SPT beam window functions $B_{\ell}$ are measured by combining dedicated observations of planets and point source signals from CMB field data \citep{henning18, dutcher21}. 
The real-space beams roughly correspond to Gaussians with $\theta_{\rm FWHM} = 1.^{\prime}7$, $1.^{\prime}2$, $1^{\prime}$ in the \firstsptband, 150, and 220 GHz bands, respectively, though we use the measured $B_{\ell}$ in this analysis.
For \herschel-SPIRE, we approximate the beams as Gaussian with $\theta_{\rm FWHM} = 36.^{\prime \prime}6$ and $25.^{\prime \prime}2$ for the 600 and 857 GHz bands \citep{viero19}.

\ifdefined\PRformat
\subsection{A.2. Post-map processing}
\else
\subsection{Post-map processing}
\label{subsubsec_post_map_processing}
\fi

We obtain the calibration factor $T_{\rm cal}$ for SPT maps by cross-correlating them with \planck{} individual frequency maps. 
Since the SPT filtering will affect this cross-correlation estimate, we mock-observe the \planck{} maps using our map-making pipeline. 
The value of $T_{\rm cal}$ depends on the frequency band and the SPT survey. 
It is obtained as $T_{\rm cal} = \dfrac{C_{\ell}^{{\rm SPT} \times \planck}}{C_{\ell}^{\rm SPT-H1 \times SPT-H2}}$
averaged in the multipole range $\ell \in [100, 1500]$. 
Note that we use SPT observations from the two halves (H1 and H2) to remove the effect of noise bias. 

We optimally combine the calibrated maps from SPT and SPIRE observations using a harmonic space linear combination technique \mbox{$S_{\ell} = \sum_{i=1}^{\rm N_{\rm bands}} w_{\ell}^{i} M_{\ell, m}^{i}$} \citep{cardoso08}. 
In this study, we work with a \finalilckeynamelong{} \finalilckeyname{} map for which the frequency-dependent weights correspond to \mbox{$w_{\ell}^{\rm MV} =  \dfrac{\clinv A_{S}}{A_{S}^{\dagger} \clinv A_{S}}$}, where ${\bf C}_{\ell}$ is the covariance matrix containing the covariance between maps in multiple frequencies at a given $\ell$ and has dimension $\rm N_{\rm bands} \times \rm N_{\rm bands}$ and \mbox{$A_{S} = [1 ... 1]$} is a $1 \times \rm N_{\rm bands}$ vector containing the frequency response vector of the CMB. 
Since the SPT noise is anisotropic, with strong features along the scan direction, we perform this combination using a 2D-LC and the weights are a function of both $\ell_{x}$ and $\ell_{y}$ modes. 
We deconvolve the beams and the TF from individual bands before the linear combination step and apply an effective beam ($B_{\ell, {\rm eff}}$) corresponding to the beam of the 150 GHz channel of the \sptthreeg{} survey ($B_{\ell}^{\rm 150, SPT-3G}$) to the \finalilckeyname{} map.


The covariance matrix ${\bf C}_{\ell} = {\bf C}_{\ell}^{\rm astro} + {\bf N}_{\ell}$  receives contribution from both astrophysical signals and the experimental noise terms. 
For the signal portion ${\bf C}_{\ell}^{\rm astro}$, we make use of the \agora{} simulations. 
We do not find a significant difference in the residuals when we switch the \agora-based covariance to data-based \citepalias[see Appendix C and Fig. C4 of][]{raghunathan23}. 
The noise power spectra ${\bf N}_{\ell}$ for different bands are obtained using sign-flip realizations \citep{dutcher21}. 

Given that the weights are applied in Fourier space, the presence of bright point sources and clusters in our maps can introduce artefacts. 
Moreover, like in CMB lensing, source masking can also lead to significant mean-field bias \citep{benoitlevy13}. 
To mitigate such effects, we use the inpainting\footnote{\text{\faGithub} \url{https://github.com/sriniraghunathan/inpainting}} technique \citep{benoitlevy13, raghunathan19c} and reconstruct the primary CMB at the location of sources and clusters. 
The radius used for inpainting depends on the flux level of the source and the size of the clusters. 
We inpaint locations of point sources detected in our maps with $\snr \gtrsim 5$ which corresponds to a flux level of $S_{150} \ge 2\ {\rm mJy}$ 
and clusters detected with $\snr \ge 4.5$. 
The $\snr$ thresholds are determined based on a matched-filtering approach to optimally extract clusters and mm-wave point sources. More details about the techniques can be found in \citet{archipley24, kornoelje24}.
The masked point sources and clusters are uncorrelated with the \reionkszname{} and, as discussed in Appendix B.1, we correct the $\clkk$ for the \mbox{$\fskylost = 0.18$} due to masking. While the masking slightly ($\sim10\%$) reduces the $\snr$, it helps in reducing the level of the foreground signals significantly. For example, in $\ell \in [\elminvalue, \elmaxvalue]$ the point source power drops by $\times 10$ when we replace the masking threshold from $S_{150} \ge 20\ {\rm mJy}$ to $S_{150} \ge 2\ {\rm mJy}$. Similarly, masking the detected ($\snr \ge 4.5$) clusters reduces the tSZ power by $\ge \times 1.5$ \citep{raghunathan22}.

\ifdefined\PRformat
\section{B. Methods}
\label{sec_methods}
\subsection{B.1 Quadratic estimator for $\bigkmap$} 
\else
\section{Methods}
\label{sec_methods}
\subsection{Quadratic estimator for $\bigkmap$}
\label{sec_qe}
\fi
To reconstruct the $\bigkmap$, we follow the work of \citetalias{smith17} and develop a quadratic estimator (QE). The QE is similar to the one used for CMB lensing reconstruction but here the estimator looks for the velocity-induced, rather than lensing-induced, correlation between the otherwise independent $\ell$ modes. 

The QE works by computing the product of two Wiener-filtered CMB temperature maps $X(\hat{\bf n})$ and $Y(\hat{\bf n})$ to construct $\bigkmap = X(\hat{\bf n}) Y(\hat{\bf n})$ or equally in harmonic space as $\bigk_{L} = \int d^{2}{\ell} X_{\ell} Y^{\ast}_{\ell-L}$.
The Wiener filter $W_{\ell}$ is employed to down-weight the modes contaminated by sources such as CMB, astrophysical foregrounds, and instrumental noise. 
As mentioned previously, we use the \finalilckeyname{} map for both $X(\hat{\bf n})$ and $Y(\hat{\bf n})$.
Given that the CMB is much brighter than the kSZ on large scales, we explicitly zero modes in our maps at $\ell < \elmin$, following \citetalias{smith17}. 
Similarly, we also remove the contribution from small scales $\ell > \elmax$ as they are dominated by foregrounds and instrumental noise. 
The values used for $\elmin$ and $\elmax$ are different from \citetalias{smith17}, though, and are based on a thorough investigation of the impact of foreground signals as discussed below. 
The combination of these bandpass filters and the Wiener filter for the \finalilckeyname{} map yields 
\begin{eqnarray}
W_{\ell} = \left\{
\begin{array}{l l}
\dfrac{C_{\ell}^{\rm kSZ}} {(C_{\ell}^{\rm kSZ} + C_{\ell}^{\rm CMB} + N_{\ell}^{\rm \finalilckeyname})}&, \ell \in [\elmin, \elmax]\\~\\
0&, ~{\rm otherwise}
\end{array}\right.
\label{eq_final_filter}
\end{eqnarray} where 
$C_{\ell}^{\rm kSZ}$ is the expected total kSZ power spectrum which we assume to be flat in $D_{\ell}^{\rm kSZ} \equiv \ell (\ell+1) C_{\ell}^{\rm kSZ}/2 \pi$ with an amplitude of $3\ \mu K^{2}$ \citep{battaglia13, reichardt21}, $C_{\ell}^{\rm CMB}$ corresponds to the lensed CMB power spectrum calculated using \texttt{CAMB} \citep{lewis00}, and $N_{\ell}^{\rm \finalilckeyname}$ is the residual noise and foreground power spectrum in the \finalilckeyname{} map. 
The noise power spectrum used for $N_{\ell}^{\rm \finalilckeyname}$ is calculated using the sign-flip noise maps from data and the foreground power spectrum is computed using \agora{} simulations.
We set $\elmin = \elminvalue$ and $\elmax = \elmaxvalue$ as the fiducial values and discuss more about this choice in \ifdefined\PRformat the Appendix. \else Appendix~\ref{sec_bias_cmb_fg}.\fi 
Replacing the flat $D_{\ell}^{\rm kSZ}$ power spectrum assumption by $D_{\ell} = A_{\rm kSZ} \left( \frac{\ell}{\ell^{\ast}} \right)^{\alpha_{\rm kSZ}}$ with amplitude $A_{\rm kSZ} = 3\ \mu {\rm K}^{2}$, pivot $\ell^{\ast} = 3000$, and slope $\alpha_{\rm kSZ} = [-1, -0.5, 0.5, 1]$ results in negligible $<2\%$ shifts in the final parameter constraints.

The desired kSZ trispectrum \mbox{$\clkk = \delta(L-L^{\prime}) \bigk_{L} \bigk_{L^{\prime}}^{\ast}$} is the power spectrum of the reconstructed $\bigkmap$ map \citepalias{smith17} after removing an estimate of the mean-field that arises due to  \ifdefined\PRformat masking. \else masking (see \S{\ref{sec_N0_meanfield_covariance_estimation}})\fi 
We debias $\clkk$ to account for the above filtering $W_{\ell}$, the effective beam $B_{\ell, {\rm eff}}$ of the \finalilckeyname{} map, TF $F_{\ell}$ and the sky fraction lost ($\fskylost$) due to masking by dividing the measured $\clkk$ by $\bar{F}^{2}$ where 
\begin{eqnarray*}
\bar{F} & = & \fskyfinal^{-4} \int \frac{ W_{\ell}^{2}\  B^2_{\ell, {\rm eff}}\ F_{\ell}\ d^{2}{\ell}}{(2\pi)^{2}} \\
& = & \fskyfinal^{-4} \int \frac{W_{\ell}^{2}B^2_{\ell, {\rm eff}}\ F_{\ell}\ (2 \pi) \ell d{\ell}}{(2\pi)^{2}} \\
& = & \fskyfinal^{-4} \int \left( \frac{\ell} {\ell} \right) \frac{W_{\ell}^{2}B^2_{\ell, {\rm eff}}\ F_{\ell}\ \ell d{\ell}}{(2\pi)}
\end{eqnarray*}    
and setting $\dfrac{d\ell}{\ell} = d{\rm ln \ell}$, we obtain 
\begin{eqnarray}    
\bar{F} & = & \fskyfinal^{-4} \int \left( \frac{\ell^{2}} {2 \pi} \right) W^{2}_{\ell}\ B^2_{\ell, {\rm eff}}\ F_{\ell}\ d {\rm ln} \ell
\end{eqnarray}    
with \mbox{$\fskyfinal = \frac{100}{4 \pi} \left( \frac{\pi}{180} \right)^{2} (1 - \fskylost)$}.
As mentioned before, besides kSZ, the measured trispectrum $\clkk$ contains contributions from CMB lensing ($\clphiphi$) and foreground ($\clfg$) signals. 
There is also contribution from the Gaussian disconnected piece arising due to chance correlations of the two CMB maps, which we represent as $\nzerobias$ to continue with the CMB lensing analogy. 
Together, the total measured $\clkk$ can be defined as
\begin{eqnarray}
\clkk & = & \clkkideal + \clkksys + \nzerobias .
\label{eq_cl_kk}
\end{eqnarray}
The second term on the right is the systematic \mbox{$\clkksys = \clphiphi + \clfg + \clphifg$} where $\clphiphi$ is the lensing power spectrum, $\clfg$ is the foreground trispectrum which consists of both unlensed and lensed foreground signals, and the $\clphifg$ is the CMB lensing-foreground bispectrum. 
The foregrounds correspond to auto- and cross-trispectra from CIB, tSZ, and radio galaxies. 
We discuss these contributions in \ifdefined\PRformat the Appendix. \else Appendix~\ref{sec_bias_cmb_fg}. \fi
In this work, we ignore higher-order bias terms like $N^{(1, KK)}_{L}, N^{(2, KK)}_{L}, N^{(3/2, KK)}_{L}$, some of which are important for CMB lensing reconstruction with current low-noise CMB datasets \citep[for definitions of each term, see][]{madhavacheril20b}.
We calculate the bias terms $\nzerobias$ and $\clkksys$ using simulations described below.

\ifdefined\PRformat
\subsection{B.2 Simulations}
\else
\subsection{Simulations}
\label{sec_sims}
\fi
As mentioned before, we use the \agora{} \citep{omori22} simulations to model the effects of astrophysical foregrounds (both lensed and unlensed) and CMB lensing. 
\agora{} is a simulation suite containing correlated multi-component extragalactic sky maps, namely lensed CMB, CIB, kSZ, radio, and tSZ. 
The components are also correlated across the different bands. 
It has been generated using the dark matter particles and halo catalogs from MultiDark Planck 2 (MDPL2, \citealt{klypin16}). 
Note that the kSZ map in \agora{} corresponds to the post-reionization kSZ signal due to halos in the local Universe and not the \reionkszname. 

To model the \reionkszname, we use the Abundance Matching Box for the Epoch of Reionization (\amber\footnote{\url{https://github.com/hytrac/amber/tree/main}}) simulation package \citep{trac22, chen23}. 
\amber{} has the capability to produce the \reionkszname{} signal as a function of the following parameters: midpoint of reionization $\zmid$; duration\footnote{In \amber, the default parameterization of the duration of reionization $\zduramber$ is the period corresponding to the difference in redshifts at which the Universe has been 5\% and 95\% reionized.} $\zduramber$; asymmetry of reionization $A_{z}$ with respect to $\zmid$; minimum halo mass $M_{\rm min}^{h}$ used for reionization; and the opacity of the ionized bubbles parameterized using mean free path $\lambda_{\rm MFP}$. 
We find that the changes in the kSZ 4-pt function are negligible when we tweak the parameters $\theta \in [A_{z}, {\rm log}(M_{\rm min}^{h}), \lambda_{\rm MFP}]$ and fix them to $A_{z} = 3$, ${\rm log}(M_{\rm min}^{h}) = 8$, and $\lambda_{\rm MFP} = 3$\citep{chen23}.
We set the boxsize to $\amberboxsize$ with $\ambernparticles$ particles corresponding to a resolution of $\sim \amberresolution$. 
We do not find a significant change in the kSZ 4-pt function signal when we increase the boxsize by a factor of $\times 1.5$ at fixed resolution.

Upon running \amber{} for different values of $\zmid$ and $\zduramber$, and averaging over 10 realizations, we find that the average kSZ 4-pt function in \mbox{$L \in [50, 300]$} can be described by 
\begin{eqnarray}
\label{eq_amber_ksz_4pt_DL}
L \clkkideal = A \left[ \dfrac{\zduramber}{\zduramber^{\ast}} \right]^{\alpha} \left[ \dfrac{\zmid}{z_{\rm re}^{\rm mid^{\ast}}} \right]^{\beta}
\end{eqnarray} 
with $A = 2.73 \times 10^{-5}$, $\alpha = 1.74$, and $\beta = 2.52$ for pivot points $\zduramber^{\ast} = 4$ and $z_{\rm re}^{\rm mid^{\ast}} = 8$. 
We convert $\zduramber$ to $\zdur$, the standard definition in literature, using the relation \mbox{$\zdur = 0.387 \zduramber + 0.009$} for \mbox{$A_{z} = 3$} \citep{chen23}. 

To test our pipeline we require multiple realizations of the \reionkszname{} and the astrophysical foreground signals. 
While the former is easier to generate using \amber, generating multiple realizations of correlated non-Gaussian foregrounds is computationally expensive. 
Instead we add the \amber-kSZ signal to the single available \agora{} full-sky map and extract \howmanysimulations{} different sky patches corresponding to the size of our field. 
Given that our field is only \fieldsize, the \howmanysimulations{} patches have no overlap, 
and we treat each of these patches as a separate sky realization. 

We also generate $\howmanysimulationsfornzero$ correlated Gaussian realizations of lensed CMB and foregrounds using the power spectra $C_{\ell}$ estimated from \agora{} simulations. 
These realizations are used to estimate the mean-field, $\nzerobias$, and the covariance. 
We also include the \reionkszname{} with the above simulations. This is done by creating Gaussian realizations using the power spectrum of the \amber{} kSZ map with $\zdur = 4$ and $\zmid = 8$.
Modifying the values of $\zdur$ and $\zmid$ has negligible impact on the above terms. 

We filter the simulations using the 2D TF calculated from mock observations. 
The filtered simulations are then convolved with the beam corresponding to each band. 
Next we add noise to the simulations obtained using the sign-flip realizations. 
The simulations are processed in the same way as the data (see Appendix) and then we construct the \finalilckeyname{} maps from both the non-Gaussian and Gaussian versions. 

\ifdefined\PRformat
\subsection{B.3 $\clkk$ estimation and Likelihood}
\else
\subsection{$\clkk$ estimation and Likelihood}
\label{sec_estimation}
\fi
We pass the \finalilckeyname{} maps from both data and the two sets of simulations through the QE pipeline to reconstruct the $\bigkmap$ maps and estimate $\clkk$. 
As described in the main text, for the likelihood calculation, we use $\clkk$ in the range $L \in [50, 300]$ with $\Delta_{L} = 50$, since beyond that the kSZ 4-pt function is expected to be sub-dominant compared to other components \citepalias{smith17}. 
We assume a Gaussian likelihood of the form
\begin{equation}
    \label{eq_likelihood}
    -2 \ln{} \mathcal{L}(\mathbf{d} | \theta) = \sum_{L L^{\prime}} \left(\clkk - \clkkideal \right) \covforlikelihoodfullinv \left(\hat{C}_{L^{\prime}}^{KK} - C_{L^{\prime}}^{KK} \right)\,,
\end{equation}
where $\clkk$ is the reconstructed 4-pt 
containing the input kSZ, CMB lensing and foregrounds after removing $\nzerobias$; $\clkkideal(\theta) \equiv \clkkideal$ is the model vector as a function of parameters $\theta \in [\zmid, \zdur]$; and $\covforlikelihoodfull$ is the covariance matrix. 
While we have assumed a Gaussian likelihood for the current work, the shape of the non-Gaussian covariance might become important for the future surveys. Subsequently, techniques such as likelihood-free or simulation-based inferences, which can model any arbitrary distribution, might be better suited.

\ifdefined\PRformat
\subsection{B.4 Mean-field, $\nzerobias$, $\clkksys$ and $\covforlikelihoodfull$}
\else
\subsubsection{Mean-field, $\nzerobias$, $\clkksys$ and $\covforlikelihoodfull$ estimation}
\label{sec_N0_meanfield_covariance_estimation}
\fi
Similar to CMB lensing reconstruction, the presence of point source masks, boundary masks, and spatially varying signals will result in a non-zero ``mean-field'' bias. 
Note that we do not use point source masks as we inpaint the locations of detected point sources and clusters, and our noise is uniform across the \fieldsize{} field. 
As a result, our mean-field should be dominated by our boundary mask.
We derive the mean-field by averaging the $\bigkmap$ maps from \howmanygaussiansimulations{} Gaussian realizations. 
This estimate is subtracted from all the $\bigkmap$ before computing $\clkk$. 
The $\clkk$ from $\howmanygaussiansimulations$ Gaussian realizations is averaged to estimate the disconnected Gaussian piece $\nzerobias$. 

We use the non-Gaussian realizations to build a template for $\clkksys$. 
Specifically, we estimate it as the mean reconstructed 4-pt function of all the \howmanysimulations{} simulations.  
Given the imperfect knowledge of foregrounds, we also estimate a systematic error on this template by scaling the tSZ/CIB signals in the input \agora{} simulations by 20\%. As we show later, these uncertainties do not have a significant impact on our results.

The covariance matrix used for likelihoods is computed using $\clkk$ measurements from the non-Gaussian realizations. 
This takes into account the scatter from both the Gaussian disconnected terms and the non-Gaussian signals. 
We also compute the covariance from the Gaussian realizations $\covforlikelihoodgau$ which we use to report the raw detection significance of the trispectrum measurement.

\ifdefined\PRformat
\subsection{B.5 Pipeline validation}
\else
\section{Pipeline validation}
\label{sec_pipeline_validation}
\fi

We use simulations to validate our pipeline and the $\clkksys$ template. 
The distribution of $\clkk$ from the simulations are shown as the violins in Fig.~\ref{fig_clkk_data_sims}. The simulations contain noise, and the the non-Gaussian CMB and foreground signals.
But for these tests we also inject a \reionkszname{} signal assuming $\zmid = \planckzmid$ and $\zdur = 4$ to the above simulations.  
We report the results using constraints on the $\zdur$ parameter. 
Since $\zmid$ and $\zdur$ are highly degenerate, for these tests we also apply a Gaussian prior on $\zmid$ centred at \mbox{$\zmid = \planckzmid \pm \planckzmiderror$}. 
This prior roughly corresponds to \planck's measurement based on the low-$\ell$ $E$-mode reionization bump. 
We do note that \planck's optical depth measurement error cannot be directly translated into a flat prior on $\zmid$ since that conversion depends on the reionization history $X_{\rm e}(z)$. 
For the simulations, however, the signal is injected assuming a fixed $\zmid = \planckzmid$ and not based on $\taure$, and as a result this flat prior on $\zmid$ should not bias the results. 
We follow two approaches to account for the potential mismatch between the $\clkksys$ from data and \agora.
\begin{itemize}
\item \mbox{Case (1):} In this approach, we remove the $\clkksys$ template computed using \agora{} from the measured $\clkk$ and
only fit for the two EoR parameters $\theta \in [\zmid, \zdur]$. 
This is optimistic in terms of $\snr$.
\item \mbox{Case (2):} In the second approach, we use the \agora{} $\clkksys$ template but rather than removing it, we fit for an additional amplitude term $\ampcmbfg$. In this case, we fit for three parameters $\theta \in [\zmid, \zdur, \ampcmbfg]$. We assume a flat prior on $\ampcmbfg \in [0, 5]$. \textbf{This is our baseline approach}.
\end{itemize}

\begin{figure*}
\centering
\includegraphics[width=0.95\textwidth, keepaspectratio]{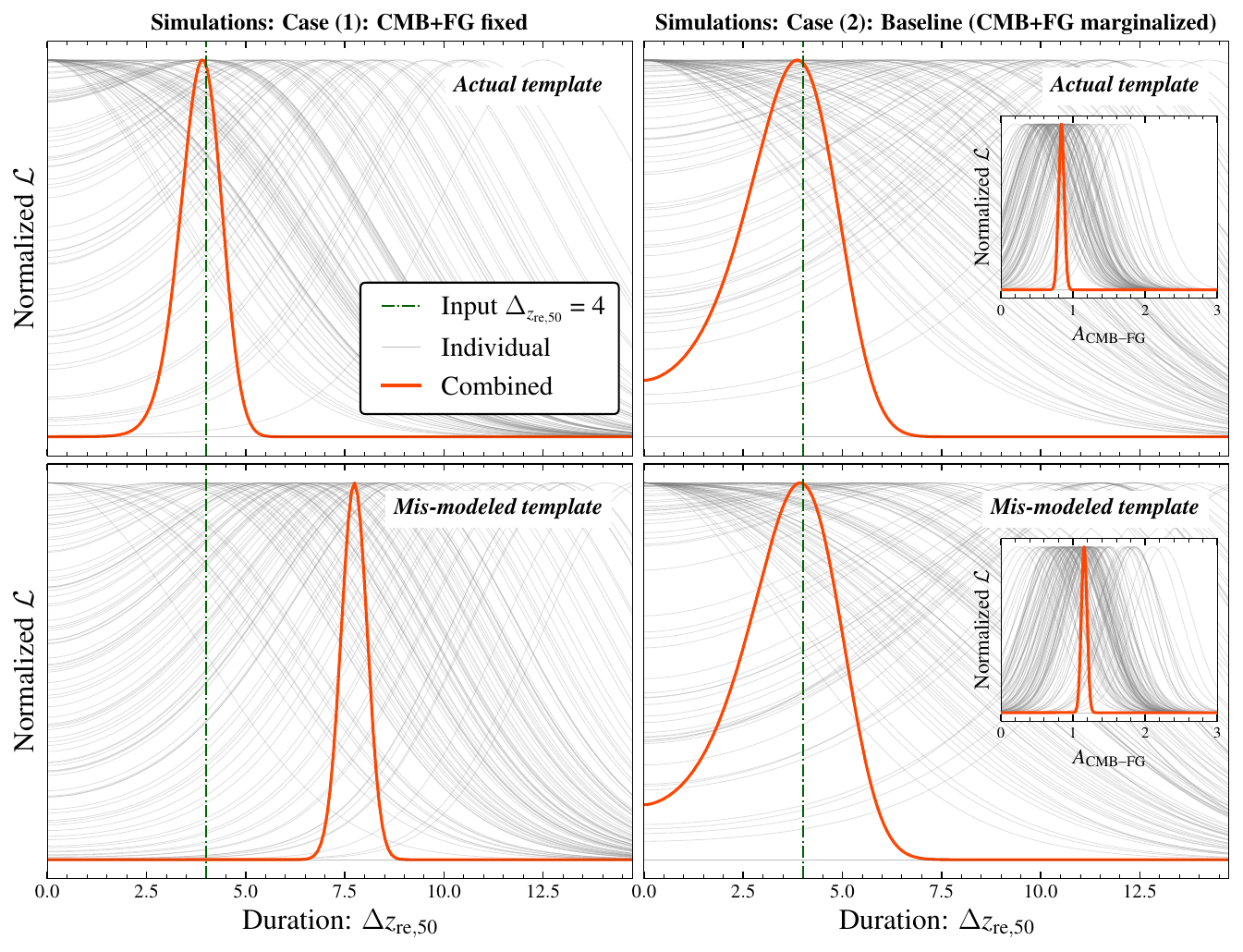}
\caption{Marginalized posterior distributions from simulations for $\zdur$ for Case (1) in the left panels and the baseline Case (2) in the right panels. 
The gray curves represent the individual simulations and the orange curve is the combined posterior from \howmanysimulations{} simulations. 
In the top panels, the template matches the input simulations (i.e:) $\clkksys$ is the mean of all the \howmanysimulations{} simulations in Fig.~\ref{fig_clkk_data_sims}. 
The bottom panels are for the case when the input simulations have the tSZ signal scaled by $\times 1.2$ but the templates used are the same as in the top panels. 
In both the top panels, when the templates match, the orange curve recovers the input $\zdur = 4$ (green dash-dotted curve) in an unbiased way. 
We also note that the width of the orange curve in the right panel is wider when $\ampcmbfg$ is marginalized compared to when it was fixed in the left panel. 
In the bottom left panel, however, when the templates are mis-modeled, the orange curve is significantly biased ($5.5 - 6 \sigma$) when $\ampcmbfg$ is fixed. 
But in the bottom right panel, we again recover unbiased results.
Also shown in the inset plots are the marginalized posterior for $\ampcmbfg$ parameter. 
We note in the bottom right, the orange curve shifts slightly to higher values compared to the top right to account for the enhanced tSZ signal in the input simulations.}
\label{fig_delta_z_posterior_sim_tests}
\end{figure*}

The results are presented in Fig.~\ref{fig_delta_z_posterior_sim_tests}. 
The individual simulations are in grey and the combined likelihood from all the $\howmanysimulations$ simulations is in orange. 
In the figure, left and right panels correspond to Case (1) and Case (2) described above. 
The simulations used in the top panels are the fiducial \howmanysimulations{} non-Gaussian simulations from Fig.~\ref{fig_clkk_data_sims} while in the bottom panels we scale the tSZ portion of the simulations by $\times1.2$. 
The templates used are always the same -- the black curve from Fig.~\ref{fig_clkk_data_sims} which is the mean of all the \howmanysimulations{} simulations --.

In the top panels, the orange curve recovers the input in both \mbox{Case (1)} and \mbox{Case (2)} with values \mbox{$\zdur = \zdurcombinedsimsbestfitcasea \pm \zdurcombinedsimserrorcasea$} and \mbox{$\zdur = \zdurcombinedsimsbestfitcaseb \pm \zdurcombinedsimserrorcaseb$}, respectively.
The error on $\sigma(\zdur)$ is higher for \mbox{Case (2)} because of the additional degree of freedom, as expected. 
In the bottom panel, however, the orange curve in the left panel is significantly biased ($5.5-6\sigma$) while the right panel, in which we fit for the amplitude of $\ampcmbfg$, returns an unbiased result. 
The inset plots in the right panels show the marginalised posteriors for $\ampcmbfg$. 
The combined posterior shifts slightly to higher values in the bottom panel compared to the top panel to account for the enhanced tSZ signal in the simulations. 
We also note that the joint likelihood from the \howmanysimulations{} simulations in the top panel results in $\ampcmbfg$ that is smaller than 1 and this could be due to the mismatch in the template shapes between different simulations. This parameter, however, is not of interest and the shift should be much smaller for a single simulation run. Moreover, this does not cause any impact on the recovered $\zdur$ and hence we do not investigate this further.

Given these results, we choose Case (2) as our baseline approach. 
For both Cases (1) and (2), the detection significance for data or equivalently a single simulation run is roughly $\zdurcombinedsimssnrcasea/\sqrt{100}$ and $\zdurcombinedsimssnrcaseb/\sqrt{100}$ which are both $<1\sigma$. 
Subsequently, we do not expect a detection of $\zdur$ and only place upper limits (95\% C.L) in this work.

Below, we discuss more about how the biases change when the CIB and tSZ signals are further modified, and also when $\ell$ ranges used for the filter in Eq.(\ref{eq_final_filter}) are modified.

\ifdefined\PRformat
\section{C. Importance of $\ell$ ranges to maximize $\snr$ and mitigate biases}
\else
\section{Importance of $\ell$ ranges to maximize $\snr$ and mitigate biases}
\label{sec_bias_cmb_fg}
\fi

The choice of $\elmin$ and $\elmax$ in Eq.(\ref{eq_final_filter}) depends on two things: the measurement $\snr$ and the systematic biases arising from CMB lensing and foregrounds. 
The value for $\elmin$ is primarily motivated to reduce the contribution from CMB to the $\nzerobias$ and $\clkksys$ estimates. 
Similarly, the value for $\elmax$ is chosen to reduce the contribution of foregrounds, which tend to increase on small scales. 
Like CMB, foregrounds also impact both the $\nzerobias$ and $\clkksys$. 
While the bias from the CMB lensing term is relatively easy to model, thanks to high-significance measurement of CMB lensing from \sptpol{} in our 100 $\sqdeg$ field \citep{wu19, millea21}, modeling the foreground 4-pt function signal is hard as it involves understanding each of the foregrounds (tSZ, CIB, and radio) individually and also the correlation between them. 

As described above, we use the \finalilckeyname{} map without explicitly projecting out any foreground components using, e.g., constrained-ILC (cILC, \citealt{remazeilles11}). 
While removing the tSZ signal can be helpful for tSZ-induced biases, the process can enhance the CIB residuals significantly \citepalias[See Fig. 3 of][]{raghunathan23}. 
Similarly, nulling the CIB signal enhances the tSZ residuals. 
The process of foreground removal also generally increases the noise in the final ILC maps. 
Another approach is to plug in different cILC in the two legs of the QE as proposed by \citetalias{raghunathan23}, although this operation also increases the noise in the reconstructed $\bigkmap$ map. 
Instead, we use different $\ell$ cuts on the \finalilckeyname{} map to minimize the foreground contamination. 

Besides the CMB lensing and foreground 4-pt function, which are both positive\footnote{Note that some of the foreground cross-correlations terms can be negative but the total foreground trispectrum is positive.}, there is also a bispectrum bias arising from the correlation of CMB lensing and foregrounds which is negative on large scales \citep[See Fig.14 of][]{vanengelen14}. 
If this negative bispectrum bias dominates our measurement, a non-detection of the kSZ 4-pt function cannot be used to set upper limits on the EoR parameters without a proper model for the bispectrum bias. 
In the absence of the negative bispectrum bias, however, foreground modeling is not necessary to set place upper limits on EoR using the measured 4-pt function signal. 

Our choice of the fiducial $\elmin = \elminvalue$ and $\elmax = \elmaxvalue$ 
is based on the above arguments. 
Reducing $\elmin$ increases the contribution due to CMB lensing. 
Although that can be modeled as stated above, it also significantly increases the $\nzerobias$ and the scatter due to sample variance of the CMB from these large scales. 
Similarly, increasing $\elmax$ increases the contribution from foregrounds. 
In the case where the CMB lensing and foregrounds can be perfectly modeled in the simulations, we do not expect any bias. 
However, given that the foregrounds modeled using \agora{} can be different from data, we estimate the biases due to imperfect modeling of the simulations as a function of different $\elmin$ and $\elmax$ values. 
Below, we quantify these biases and the impact on $\snr$ when the filter $\ell$ ranges are modified.

\ifdefined\PRformat
\subsection{C.1. Impact on $\snr$}
\else
\subsection{Impact on $\snr$}
\fi
Here we discuss the changes to $\snr$ when we modify our fiducial values for \mbox{$(\elmin, \elmax) = (\elminvalue, \elmaxvalue)$} and $\Delta_{\ell} = 1000$. 
We limit this test to Case (1) where the CMB+FG is fixed. 
When we reduce $\Delta_{\ell}$ from 1000 to 700 (500) keeping the $\elmin$ and $\elmax$ values fixed, the best-fit $\zdur$ from the combined $\howmanysimulations$ simulations is $\zdurcombinedsimsellchoiceaunbiaseddeltaellsevenhun \pm \zdurcombinedsimsellchoiceaunbiaseddeltaellsevenhunerr$ ($\zdurcombinedsimsellchoiceaunbiaseddeltaellfivehun \pm \zdurcombinedsimsellchoiceaunbiaseddeltaellfivehunerr$) compared to $\zdurcombinedsimsellchoiceaunbiased \pm \zdurcombinedsimsellchoiceaunbiasederr$ using $\Delta_{\ell} = 1000$, implying a reduction in the $\snr$ by roughly $20\%$ ($40\%$) for the reduced $\Delta_{\ell}$ values. 
This is not surprising given the reduced number of modes used for $\bigkmap$ reconstruction.
Similarly, we also checked the effects of sliding the ($\elmin, \elmax$) values lower or higher keeping the $\Delta_{\ell} = 1000$ fixed. 
Moving them to lower values (large scales) with $(\elmin, \elmax) = (3000, 4000)$, reduces the $\snr$ by $14\%$ due to the increased sample variance of CMB.
Sliding the filters to higher values with $(\elmin, \elmax) = $ (3700, 4700) and (3500, 4500) also reduces the $\snr$ by 5\% and 14\% respectively due to the increased scatter from foregrounds and noise on small scales.

\ifdefined\PRformat
\subsection{C.2. Bias due to mismatch in $\clkksys$ template} 
\else
\subsection{Bias due to mismatch in $\clkksys$ template} 
\label{sec_bias_template_mismatch}
\fi
Now, we discuss the bias in the recovered $\zdur$ from the mismatch between the true CMB+FG 4-pt function $\clkksys$ template and the one estimated using \agora.
We perform this test for both 
Case (1)
and Case (2) using simulations where we modify the tSZ or the CIB signals in the input simulations by $\times1.2$. 
The $\clkksys$ template used for the fitting process, however, does not include the tSZ or CIB scaling. 

For the combined \howmanysimulations{} simulations in Case (1), we see a $\sim 5.5-6\sigma$ shift due to the mismatch in the tSZ signal used for input and fitting (see bottom left panel of Fig.~\ref{fig_delta_z_posterior_sim_tests}). 
While this is large, note that the bias in a single simulation run is much smaller at $\le 0.5\sigma$.
Sliding the filters to larger scales $(\elmin, \elmax) = (3000, 4000)$ results in a slightly smaller (2\%) level of bias. 
Moving the $(\elmin, \elmax)$ values of the filters to higher values increases the bias with roughly $0.73\sigma$ for $(\elmin, \elmax) = (3700, 4700)$. 
The biases are $\sim 10\%$ higher when the CIB signal is scaled by $\times 1.2$ rather than the tSZ.
When we reduce the scaling to $\times1.1$, the bias goes down, as expected.

For Case (2), when we fit for the amplitude of $\ampcmbfg$, however, we do not observe any bias (see bottom right panel of Fig.~\ref{fig_delta_z_posterior_sim_tests}). 
When we use the baseline $\clkksys$ template without tSZ scaling to analyse the results from simulations where the tSZ is enhanced by $\times1.2$, we recover $\zdur = \zdurcombinedsimscasebfortszoneptwosimcase \pm \zdurcombinedsimserrorcasebfortszoneptwpsimcase$. 
At this juncture, we would like to emphasise that we do not simply mismatch the $\clkksys$ template by a simple amplitude scaling but instead modify the tSZ or CIB signal to generate $\clkksys$. This process also changes the shape of the $\clkksys$ template. 

Based on these results, we claim our baseline approach Case (2) is not affected by the mismatch in the shape of the template used.


\ifdefined\PRformat
\subsection{C.3. Bispectrum bias from CMB lensing and foregrounds}
\else
\subsection{Bispectrum bias from CMB lensing and foregrounds}
\fi
While the bias due to tSZ and CIB is 2\% smaller when sliding the filters towards lower $\ell$ for $(\elmin, \elmax) = (3000, 4000)$ compared to the fiducial $(3300, 4300)$, as described above, the $\snr$ goes down for the former by $14\%$. 
Another important factor to be considered when sliding the filters to larger scales is the bispectrum bias due to the correlation between CMB lensing and foregrounds \citep{vanengelen14}.
We estimate this by running two sets of non-Gaussian simulations. 
The first case is our baseline \howmanysimulations{} simulations from \agora{} where the lensed CMB and foreground signals are correlated. 
For the second case, we randomize the lensed CMB signals, thus removing the correlation between CMB lensing and foregrounds. 
Next we compare the $\clkk$ measured from the mean of the \howmanysimulations{} simulations using the two sets. 
The CMB lensing and foreground 4-pt function signals should be the same in both the sets but the second set should not have the bispectrum bias.
Subsequently, when we subtract the latter from the former, we should obtain negative signals if the bispectrum bias dominates the measurement. 
For all our filter choices $\elmin \in [3000, 3300, 3500, 3700]$ with $\Delta_{\ell} = 1000$, the difference between the two sets are consistent with a null signal indicating that the bispectrum bias is negligible.

Based on the above arguments, our fiducial choice of $(\elmin, \elmax) = (\elminvalue, \elmaxvalue)$ is an optimal choice in terms of both the $\snr$ and biases. 
While it is possible to set $\Delta_{\ell} > 1000$ retaining the fiducial $\elmin = \elminvalue$, we do not do so because of the potential issues due to foregrounds on smaller scales. 

\ifdefined\PRformat
\section{D. Other systematic checks}
\else
\section{Other systematic checks}
\label{sec_other_sys_checks}
\fi

We now estimate the systematics due to other sources namely the uncertainties in beam and TF. 
As mentioned previously, the \sptthreeg{} beams are obtained by combining measurements of planets and point sources in our field. 
While planets have a higher $\snr$, we find issues due to detector nonlinearities near the peak response and we use stacked observations of point sources in those regions. 
Modifying the radius where these two observations are combined results in differences in beam $B_{\ell}$ which is close to $2\%$ for 90 and 150 GHz bands in our fiducial $(\elmin, \elmax) = (\elminvalue, \elmaxvalue)$ range. 
In the same spirit, we also modified \herschel-SPIRE beam by assuming conservative errors of $5\%$ and $10\%$. 
This alters the $B_{\ell}$ in the desired $\ell \in [3300, 4300]$ by $\le 1\%$. We also introduced beam ellipticity $\epsilon$ as $1 - \epsilon = a/b$ where $a$  and $b$ are the semi-major and semi-minor axes. We use the following values for ellipticity: $\epsilon \in [0.1, 0.15]$. All of these are conservative assumptions and consistent with the numbers reported by SPIRE\footnote{\url{https://www.cosmos.esa.int/web/herschel/spire-overview}} experiment in \citet{swinyard10}.
Similarly, for the simulations used in this work, we have approximated the TF for all SPT bands to the 150 GHz \sptthreeg{} TF. 
In the desired $\ell$ range, we find that there are $\sim 2-5\%$ differences in TF (in power units). 
We check the impact of these using simulations and find negligible shifts ($0.05-0.2\sigma$) in the recovered $\zdur$ values. 

\ifdefined\PRformat
\section{E. Constraints on EoR and CMB+FG template from kSZ 4-pt}
\else
\section{Constraints on EoR and CMB+FG template from kSZ 4-pt}
\label{sec_eor_fg_constraints_ksz4pt}
\fi
\begin{figure*}
\centering
\includegraphics[width=0.8\textwidth, keepaspectratio]{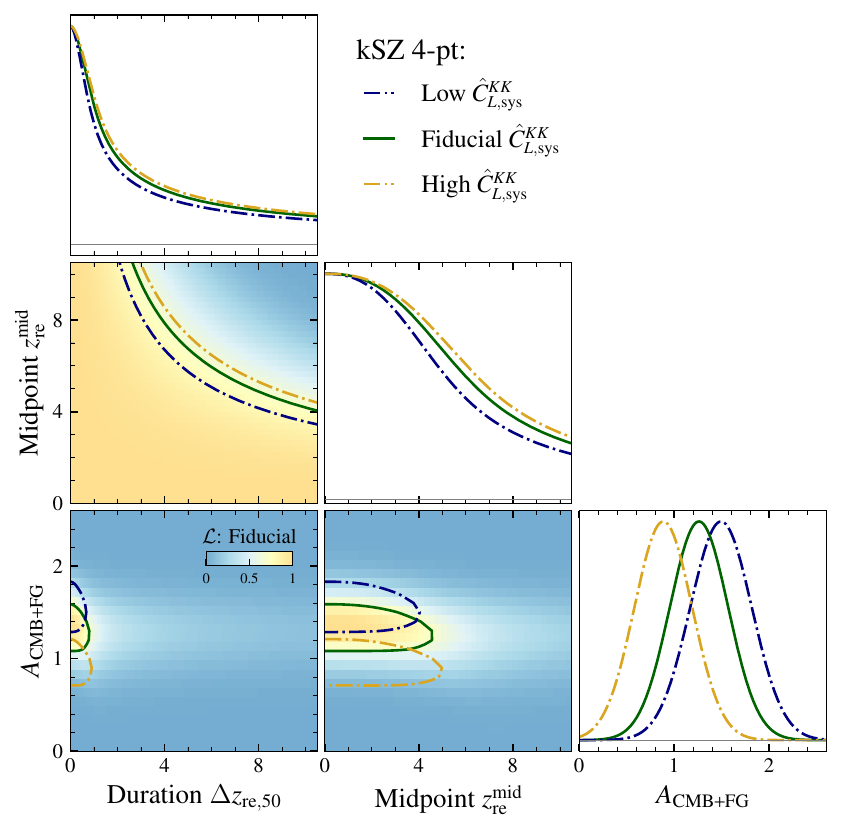}
\caption{
Constraints on EoR ($\zdur$ and $\zmid$) and $\ampcmbfg$ parameters from kSZ 4-pt only. 
Similar to Fig.~\ref{fig_posterior_data}, we present both the 1d and 2D posteriors, and the curves in the 2D plots correspond to the 68\% C.L. allowed regions.
We do not observe a strong degeneracy between $\ampcmbfg$ and EoR parameters.  
The green curves correspond to the results with the fiducial $\clkksys$ while blue (yellow) correspond to low (high) $\clkksys$ after scaling the tSZ signal down (up) by 20\% in the \agora{} simulations. 
}
\label{fig_posterior_data_ksz_4pt_only}
\end{figure*}

Here we present the constraints on both EoR and $\ampcmbfg$ parameters, in particular to argue that they are not degenerate. 
Given this is the case, we do not report the constraints on $\ampcmbfg$ after including the GP-based prior. 
This is also evident from the differences in shapes between the \amber{} kSZ 4-pt (in orange shade) and the $\clkksys$ (black) in Fig.~\ref{fig_clkk_data_sims}. 
While it is true that we could better constraints on $\ampcmbfg$ by including higher multipoles $L$, given that the degeneracy between $\zmid, \zdur$ and $\ampcmbfg$ is not strong and the fact that kSZ 4-pt is small compared to foregrounds, we do not extend the multipole range for parameter constraints. 
We present constraints for three cases. 
Green curves for the fiducial $\clkksys$ template and in this case we obtain $\ampcmbfg = \ampcmbfgbestfitvaluefromdata$. 
The blue (yellow) curves are when we scale the tSZ signal low (high) by 20\% in the \agora{} simulations during the template construction. 
As expected, switching to a low (high) $\clkksys$ results in higher (lower) values of $\ampcmbfg = \ampcmbfgbestfitvaluefromdatalowsys\ (\ampcmbfgbestfitvaluefromdatahighsys)$.
\bibliography{kSZ_4pt}
\ifdefined\PRformat
\bibliographystyle{apsrev4-1}
\fi
\end{document}